\documentclass[prx,notitlepage,citeautoscript,floatfix,twocolumn,amsmath,amssymb,superscriptaddress]{revtex4-1}%

\usepackage{graphicx}
\usepackage{dcolumn}
\usepackage{bm}
\usepackage{times}

\begin{document}


\title{Colloidal particle adsorption at water/water interfaces 
with ultra-low interfacial tension
}

\author{Louis Keal}
\affiliation{ESPCI Paris, PSL Research University, Sciences et Ing\'{e}nierie de la Mati\`{e}re Molle (SIMM), CNRS UMR 7615, 75231 Paris, France}

\author{Carlos E. Colosqui}
\email{carlos.colosqui@stonybrook.edu}
\affiliation{Department of Mechanical Engineering, Stony Brook University, Stony Brook, NY 11794, USA}
\affiliation{Department of Applied Mathematics \& Statistics, Stony Brook University, Stony Brook, NY 11794, USA}

\author{Hans Tromp}
\email{hans.tromp@nizo.com}
\affiliation{Van't Hoff Laboratory for Physical and Colloid Chemistry, Utrecht University, 3584 \!CH \!Utrecht, The Netherlands}
\affiliation{NIZO Food Research, Kernhemseweg 2, 6718 ZB Ede, The Netherlands}

\author{C\'{e}cile Monteux}
\email{cecile.monteux@espci.fr}
\affiliation{ESPCI Paris, PSL Research University, Sciences et Ing\'{e}nierie de la Mati\`{e}re Molle (SIMM), CNRS UMR 7615, 75231 Paris, France}

%
%
%
%
%
\begin{abstract}
Using fluorescence microscopy we study the adsorption of single latex microparticles at a water/water interface between demixing aqueous solutions of polymers, generally known as a water-in-water emulsion.
Similar microparticles at the interface between molecular liquids have exhibited an extremely slow relaxation preventing the observation of expected equilibrium states.  
This phenomenon has been attributed to ``long-lived'' metastable states caused by significant energy barriers $\Delta{\cal F}\sim \gamma A_d\gg k_B T$ induced by high interfacial tension ($\gamma \sim 10^{-2}$ N/m) and nanoscale surface defects with characteristic areas $A_d \simeq$ 10--30 nm$^2$.
For the studied water/water interface with ultra-low surface tension ($\gamma \sim 10^{-4}$ N/m) we are able to characterize the entire adsorption process and observe equilibrium states prescribed by a single equilibrium contact angle independent of the particle size. 
Notably, we observe crossovers from fast initial dynamics to slower kinetic regimes analytically predicted for large surface defects ($A_d \simeq$ 500 nm$^2$).
Moreover, particle trajectories reveal a  position-independent damping coefficient that is unexpected given the large viscosity contrast between phases.
These observations are attributed to the remarkably diffuse nature of the water/water interface and the adsorption and entanglement of polymer chains in the semidilute solutions.  
This work offers some first insights on the adsorption dynamics/kinetics of microparticles at water/water interfaces in bio-colloidal systems.
\end{abstract}
%
%
%
\maketitle

{\let\newpage\relax\maketitle}
%
%
%
\section{Introduction}
Particles at liquid-fluid interfaces can stabilize colloidal materials (e.g., emulsions, foams) relevant to applications in materials, biomedical, and food science \cite{ramsden1903,pickering1907,sacanna2007,dickinson2010,ngai2014,zanini2017,binks2017}. 
In this work, we study the adsorption of microparticles at so-called water/water (W/W) interfaces that form between demixing aqueous solutions of polymers, known as water-in-water emulsions \cite{grinberg1997,capron2001,scholten2002}. 
Due to the ultra-low surface tension $\gamma = {\cal O}$(1-100 $\mu$N/m) of W/W interfaces, surfactants and macromolecules have small binding energies and cannot stabilize effectively water-in-water emulsions. 
Colloidal particles, however, can have significant binding energies at W/W interfaces and have emerged as an alternative to stabilize water-in-water emulsions \cite{firoozmand2009,nguyen2013,gonzalez2016,nicolai2017,Stebe2017}. 
Hence, understanding the adsorption/desorption dynamics of colloidal particles at W/W interfaces is crucial to active technologies in biology and food science for the production of stable water-in-water emulsions. 
From a fundamental viewpoint, W/W interfaces are advantageous for studying phenomena involving capillary forces and thermal interfacial motion, as they occur over larger time and length scales than for molecular liquid interfaces. 

Using confocal microscopy we are able to record the entire adsorption process for latex microspheres ($R=$ 0.5 to 6 $\mu$m) at a W/W interface between demixing polymer solutions. 
We observe an initial exponential decay to equilibrium with a position-independent damping coefficient that cannot be explained by available models for dissipative effects induced by the interface. 
Furthermore, we observe the crossover to a late thermally activated relaxation that has been theoretically predicted \cite{colosqui2013} and partially observed by Kaz et al. for microparticles at water-oil interfaces with high surface tension $\gamma = {\cal O}$(10 mN/m) \cite{kaz2012}. 
Considering the ultra-low surface tension of the W/W interface, the observed regime crossover and slow relaxation rates indicate the presence of energy barriers that are unexpectedly large.
Experimental and theoretical analyses indicate that these findings can be attributed to polymer adsorption at the particle surface and the remarkably diffuse nature of W/W interfaces.
 
%
%
\section{Materials \& Methods}
For the experiments in this work, the demixing polymer solutions were composed of dextran (average molar mass 150 kDa from Leuconostoc bacteria, purchased from Sigma Aldrich) and cold water fish gelatin (non gelling at room temperature, from Norland Products, provided by Fibfoods, Harderwijk, The Netherlands) at native pH 5.8 and salinity corresponding to 25 mM of NaCl. 
Gelatin and dextran mixed in water at proper concentration spontaneously demix forming dextran  and a gelatin  phases with well characterized phase equilibria \cite{vis2014,vis2015charge,vis2015}. 
The system studied has 10/10 \%w/w dextran/gelatin and has equilibrium concentrations 17/0.1 and 1.2/22 for the dextran  and gelatin  phase, respectively, and a tie line length $TTL= 27$~\%w/w. 
The interfacial tension of the system is determined by the empirical relation $\gamma=10^{-2.7} TTL^{3.3}$ \cite{vis2015charge} where $TTL$ is expressed in \%w/w and $\gamma$ in $\mu$N/m. 
The studied water-in-water emulsion has been extensively characterized in prior works \cite{vis2014,vis2015charge,vis2015,keal2016thesis}, relevant physical properties are reported in Table 1. 
The shear viscosity of each phase (see Table 1) determined by rheometric measurements in cone and parallel plate configurations showed no variation for shear rates of 1 to 100 s$^{-1}$. 

\begin{table}[h]
\caption{Water-in-water emulsion and W/W interface properties.}
\begin{center}
\vskip -5pt
\begin{tabular*}{0.47\textwidth}{@{\extracolsep{\fill}}lc}
\hline
Polymer composition &	10/10 \% w/w\\
Critical concentration $c^*$ & 3.5$\pm$0.2\%\\
Radius of gyration $R_g$ (dilute) & 20$\pm$5 nm\\	
Correlation length\footnote{$\xi= 0.43 R_g(c/c^*)^{-0.75}$ \cite{degennesscaling,broseta1986}} $\xi$ & 1.2 nm\\
Interfacial tension $\gamma$	& 104$\pm$5 $\mu$N/m\\
Viscosity dextran  phase	& 0.047$\pm$0.005	Pa s\\
Viscosity gelatin  phase	& 0.515 Pa s\\
\hline
\end{tabular*}
\end{center}
\vskip -10 pt
\end{table}

\begin{figure}[h!]
\center
\includegraphics[angle=0,width=0.9\linewidth]{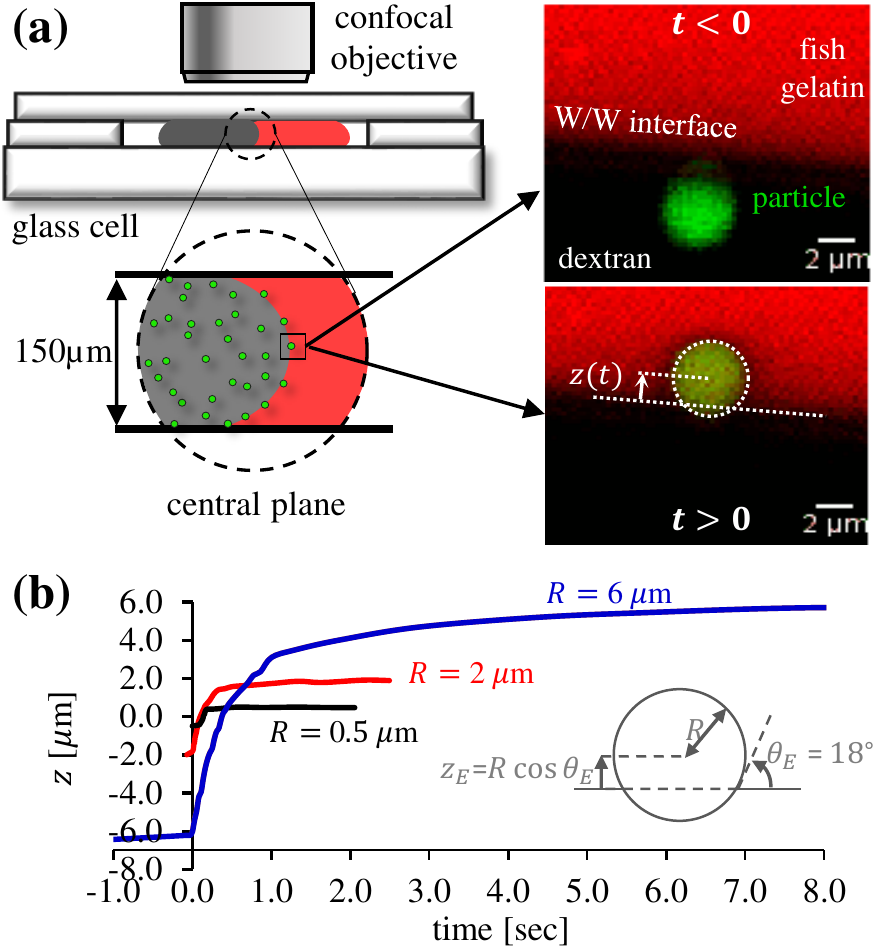}
\vskip -10pt
\caption{Experimental configuration.
(a) Left panel: glass cell (height 150 $\mu$m) confining (10/10) dextran/gelatin solutions; latex particles dispersed on the dextran phase.
Right panels: fluorescence microscope images for a 2 $\mu$m radius particle straddling the W/W interface. 
(b) Sample particle trajectories $z(t)$ for different particle radius obtained via digital processing of fluorescence microscope images. 
}
\end{figure}

To study the adsorption of single particles we produce a W/W interface with controlled curvature by bringing into contact two drops dextran and gelatin confined in a sealed chamber (cf. Fig. 1) suitable for confocal microscopy. 
The phases are obtained through centrifugation of demixed samples, after which sulfate latex particles (IDC) of the same size ($R=$ 0.5 to 6 $\mu$m) is dispersed in the dextran phase.
The curvature radius of the W/W interface is comparable to the height of the observation chamber (150 $\mu$m) and it is one to two orders of magnitude larger than the particle radius. 
In order to optically detect the W/W interface via fluorescence microscopy, rhodamine B ($\sim$0.01\%) is added to the fish gelatin phase; rhodamine B (red) stains only the gelatin phase (see Fig. 1a). 
As shown in Fig. 1, the particles are labeled with FITC (green) and thus can be easily distinguished from the gelatin phase (red) and the dextran phase (dark). 
A confocal laser-scanning microscope is adjusted to the central plane of the chamber to observe the particles that diffuse to the interface and are subsequently adsorbed (see Fig. 1a). 

The position of single particles is tracked via digital processing of image sequences (cf. Fig. 1a-b) using the open-source software ImageJ \cite{schneider2012}. 
The interface position is also tracked in order to subtract any sample drift from the particle trajectory.
The distance $z(t)$ measured normal to the W/W interface (cf. Fig. 1a) is thus recorded for several seconds (cf. Fig. 1b) using digital video at 36 frames per second, which gives a time resolution of 27.8 ms. 
The spontaneous adsorption of particles begins shortly after a particle in the dextran  phase comes in contact with the W/W interface at $z(0)=-R$ and equilibrium is reached within a few seconds for all studied particles (cf. Figure 1b). 
The equilibrium position $z_E= R \cos \theta_E$ observed for the studied particles corresponds to a size-independent contact angle $\theta_E=18^\circ$ measured through the gelatin  phase (cf. Figure 1b).

%
%
%
%
\section{Theoretical Model}
Following prior work by Colosqui and coworkers \cite{colosqui2013,rahmani2016,colosqui2016}, we assume the studied particles undergo overdamped uncorrelated Brownian motion and thus a Langevin equation
$f_d\dot{z}=-\partial{\cal F}/\partial z+\sqrt{2 k_B T f_d} \eta(t)$ 
describes the (center-of-mass) distance to the W/W interface located at $z=0$ (cf. Fig. 1).
Here, $f_d$ is the effective damping coefficient, ${\cal F}$ is the system free energy, $k_B T$ is the thermal energy, and $\eta$ is a zero-mean unit-variance Gaussian noise. 
For a spherical particle of radius $R$ and neglecting the curvature of the W/W interface, the free energy for $|z|\le R$ is approximately given by \cite{colosqui2013,rahmani2016}
\begin{equation}
{\cal F}(z)=\gamma \pi (z-z_E)^2
+\frac{1}{2} \Delta {\cal F} \sin\left(\frac{2\pi}{\ell}(z-z_E)\right)+C,
\label{eq:energy} 
\end{equation}
where $z_E=R \cos\theta_E$ is the equilibrium position determined by the equilibrium contact angle $\theta_E$ (measured in the gelatin phase), $\gamma$ is the W/W interface surface tension, and $C$ is an additive constant. 
As suggested in prior work \cite{colosqui2013}, Eq.~\ref{eq:energy} includes spatial energy fluctuations with period $\ell=A_d/2\pi R$ and amplitude $\Delta {\cal F}\sim \gamma A_d$ that are induced by nanoscale surface defects with a characteristic area $A_d\sim{\cal O}($1--10$\, \mathrm{nm}^2)$ projected on the particle surface.
Under the studied conditions we neglect terms due to line tension in Eq.~\ref{eq:energy} \cite{rahmani2016}.

%
For the case of constant damping $f_d=$~const., the initial average trajectory predicted by the proposed Langevin equation \cite{colosqui2013} is an exponential decay to equilibrium
\begin{equation}
\langle z(t) \rangle=z_E-(R+z_E) \exp(-t/T_D),
\label{eq:exp}
\end{equation}
where $T_D=f_d/2\pi\gamma$ is the characteristic decay time.
Sufficiently close to equilibrium there is a crossover to a ``slow'' kinetic regime where the average particle trajectory is nearly logarithmic in time \cite{colosqui2013}, according to
\begin{equation}
\left< z(t) \right>= z_E+
L_K \log \left[ \frac{1+A \exp(-t/T_K)}{1-A \exp(-t/T_K)}\right],
\label{eq:log}
\end{equation}
where $L_K=k_B T/ \pi \gamma \ell$ is a characteristic kinetic length, 
$A=\tanh((z_0-z_E)/2 L_K)$ is determined by the initial particle position $z_0=-R$,
and $T_K$ is the characteristic kinetic time.
The characteristic kinetic time in Eq.~\ref{eq:log} can be estimated using Kramers' rate theory \cite{colosqui2013}, which gives  
\begin{equation}
T_K= T_D 
\left(\frac{L_K}{\ell}\right) 
\frac{2\pi}{\sqrt{|\Phi^2-1|}}
\exp\left(\frac{\Delta {\cal F}}{k_BT}+\frac{1}{4}\frac{\ell}{L_K}\right),
\label{eq:time}
\end{equation}
where $\Phi= \Delta {\cal F}/4\pi\gamma \ell^2$.  
For $|\langle z \rangle-z_E|\gg L_K$, Eq.~\ref{eq:log} becomes a logarithmic trajectory
$\left<z\right>=z_E+
L_K \log(t/2T_K+c)$, where $c=\exp[-(R+z_E)/L_K]\sim 0$ is negligible small for the studied experimental conditions.
The regime crossover from capillary-driven dynamics described by Eq.~\ref{eq:exp} to a kinetic regime described by Eq.~\ref{eq:log} is a rather gradual process over a finite interval of particle positions. 
From the condition for having local minima where $\partial {\cal F}/\partial z=0$ and thus metastable states, the crossover must take place over the range\cite{colosqui2013,rahmani2016,colosqui2016}
\begin{equation}
|\left<z\right>-z_E|=\alpha \pi R \frac{\Delta{\cal F} }{\gamma A_d}
\label{eq:crossover}
\end{equation}
with $\alpha\lesssim 1$. 
Typical values $\alpha=$~0.25--0.5 in Eq.~\ref{eq:crossover} have accurately predicted the regime crossover observed in Langevin dynamics simulations and experimental studies of microparticles at a water-oil interface.\cite{colosqui2013,rahmani2016} 
%

%
\section{Results}
\begin{figure}[t!]
\center
\includegraphics[angle=0, width=0.9\linewidth]{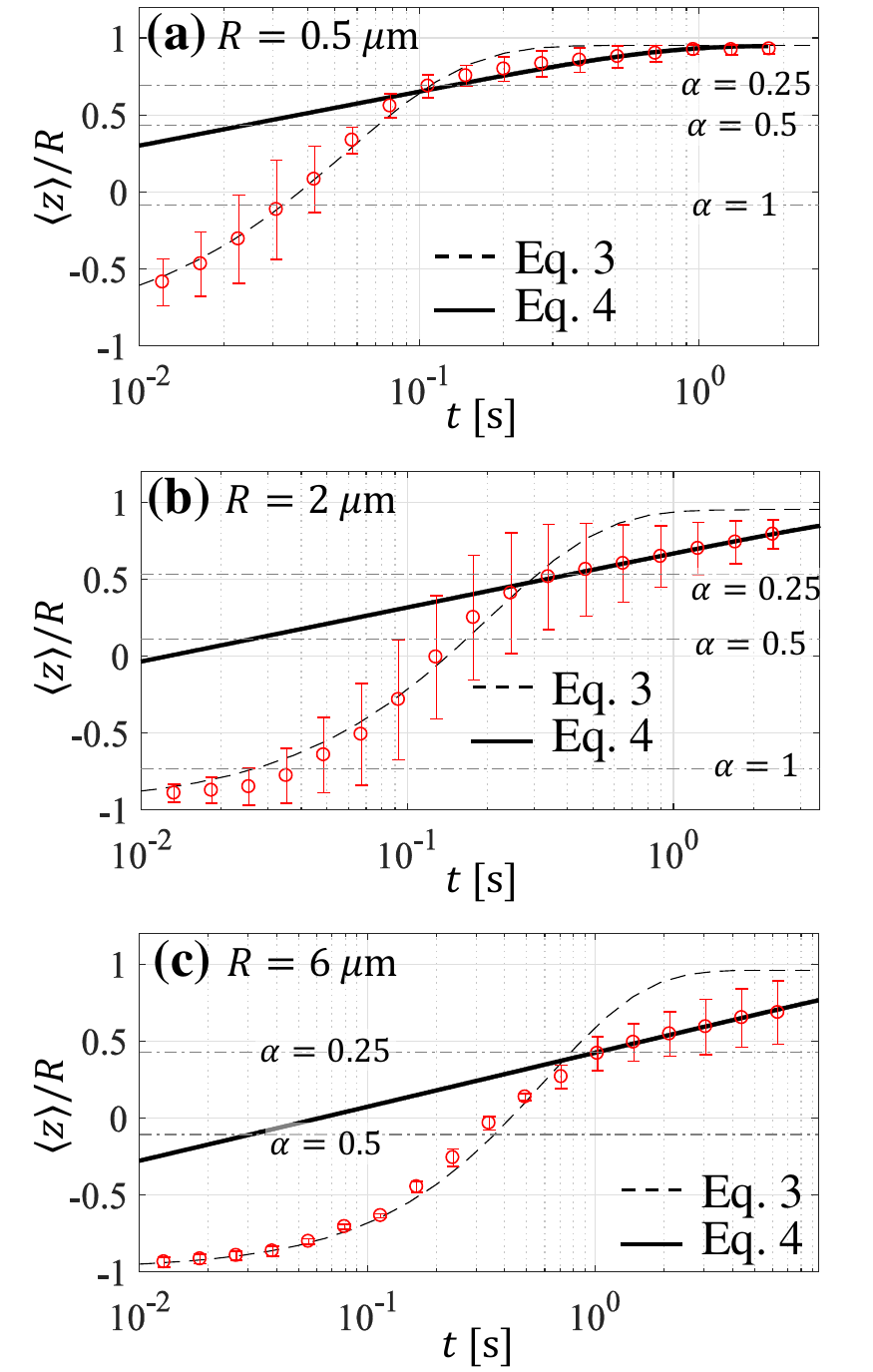}
\vskip -10pt
\caption{Average particle trajectories for single latex particles adsorbed at a W/W interface. 
(a) $R=$ 6 $\mu$m. (b) $R=$ 2 $\mu$m, (c) $R=$ 6 $\mu$m.
Experimental data (markers): average and standard deviation (error bars) obtained from 4 to 5 individual trajectories. 
Analytical fits: Eq. 2 (dashed line) and Eq. 3 (solid line) using parameters in Table 2.
The regime crossover (dot-dash horizontal lines) predicted by Eq. 5 for $\alpha=$ 0.25, 0.5, 1.
}
\label{fig:2}
\end{figure}

Fig.~\ref{fig:2} reports average trajectories $\langle z(t) \rangle$ obtained from four to five individual trajectories of particles with the same radius. 
The average particle trajectories exhibit an early exponential decay to equilibrium described by Eq.~\ref{eq:exp} and a late nearly logarithmic relaxation described by Eq.~\ref{eq:log}. 
The crossover point between these two different regimes can be predicted by Eq.~\ref{eq:crossover} using $\alpha\simeq 0.25$ (cf. Fig 3).
The decay times $T_D(R) \simeq$ 55.4, 201.4, 604.2 ms for $R=$ 0.5, 2,~\& 6 $\mu$m, respectively, were used in Eq.~\ref{eq:exp} to fit the experimental data (See Table 2) in the initial regime.  
The observed nearly exponential regime with the reported decay times corresponds to a position-independent damping coefficient $f_d=2\pi\gamma T_D=6\pi\mu_e R$, which obeys the Stokes drag formula for a spherical particle and effective viscosity $\mu_e=3.35\pm 0.05$ Pa about six times larger than the measured shear viscosity of the gelatin phase. 
A position-independent damping coefficient that scales linearly with the particle radius is consistent with scaling laws for semidilute polymer solutions originally proposed by De Gennes and co-workers \cite{degennesscaling,langevin1978,yu2002} and suggests that topological hindrance due to entanglement and polymer matrix relaxation dominate over viscous hydrodynamic effects, as elaborated in the final discussion. 
Available models for damping induced by contact lines \cite{voinov1976,cox1986,blake1969,boniello2015} are not able to account for the observed nearly exponential trajectories and linear scaling $f_d \propto R$ (see Supplemental Material). 

The decay times $T_D$ for the initial exponential trajectories in Eq.~\ref{eq:exp} are employed in Eq.~\ref{eq:time} to estimate the kinetic times $T_K=$ 0.35 to 16.5 s (see Table 2), which characterize the much slower relaxation in the thermally activated regime.
Analytical predictions from Eq.~\ref{eq:log} for the late thermally activated regime (cf. Figure 3) employ the parameters reported in Table 2. 
Notably, a single projected defect area $A_d=515$ nm$^2$, which corresponds to a rather large defect size $s_d\simeq \sqrt{A_d}$= 22.7 nm, can be employed for all the studied particles radii.
Such defect area accounts for the period 
$\ell=A_d/2\pi R=$ 0.014 to 0.16 nm  
of metastable states modeled in Eq.~\ref{eq:energy}, and thus the kinetic lengths 
$L_K=k_B T/\pi \gamma \ell=$0.08-917 nm
prescribing the displacement rates in Eq.~\ref{eq:log}. 
Energy barrier magnitudes $\Delta{\cal F}=\gamma \beta^2 A_d=$ 4.32-8.9 $k_B T$, which correspond to shape factors $\beta=$ 0.57-0.82 (see Table 2), quantitatively predict the relaxation times $T_K$ (Eq.~\ref{eq:time}) in the ``slow'' kinetic regime.  
Furthermore, the defect area $A_d$ and energy barriers $\Delta{\cal F}$ used in Eqs.~\ref{eq:log}--\ref{eq:time} predict the observed dynamic-to-kinetic regime crossover (See Fig.~\ref{fig:2}) through Eq.~\ref{eq:crossover} with $\alpha\simeq 0.25$, as predicted by Langevin simulations \cite{colosqui2013,rahmani2016}. 
These observations support the hypothesized emergence of metastable modeled in Eq.~\ref{eq:energy} states at a critical distance from the expected equilibrium position.

\begin{table}
\caption{Model parameters for analytical fits}
\begin{center}
\vskip -7pt
\begin{tabular*}{0.45\textwidth}{@{\extracolsep{\fill}}lccc}
\hline
$R$ [$\mu$m] & 0.5 & 2.0 & 6.0\\
\hline
$T_D$ [ms]	& 55.38	& 201.39 &	604.16\\
$A_d$ [nm$^2$] &	515 &	515 &	515\\
$\Delta{\cal F}/k_B T$	& 4.32 & 7.00 & 8.90\\
$L_K$ [nm] & 0.084 & 305.6 &	916.9\\
$T_K$ [s] & 0.35 & 3.15 & 16.52\\
\end{tabular*}
\end{center}
\end{table}

%

The adsorption of latex microparticles at the studied W/W interface exhibits fast exponential and slow nearly logarithmic regimes that can be quantitatively described by analytical solutions of the proposed Langevin equation for the particle position \cite{colosqui2013}.
Analytical predictions for the average trajectory $\langle z(t) \rangle$ (Eqs.~\ref{eq:exp}--\ref{eq:crossover}) give close agreement with experimental observations for the studied microparticles ($R=$ 0.5--6 $\mu$m) when using a projected defect area 
$A_d = 515$ nm$^2$ and energy barriers 
$\Delta {\cal F}=$ 4.3--8.9 $k_B T$. 
The energy barriers $\Delta {\cal F}=\gamma \lambda^2$ correspond to effective defect dimensions $\lambda=\beta\sqrt{A_d}=$~13--18 nm that are comparable to the projected size $s_d=\sqrt{A_d}\simeq 22.7$~nm. 
The effective defect size $\lambda=\beta s_d$, through the ``shape'' factor $\beta$, accounts for morphological and physicochemical properties determining the energy barriers magnitude.

\begin{figure}[t!]
\center
\includegraphics[angle=0, width=0.95\linewidth]{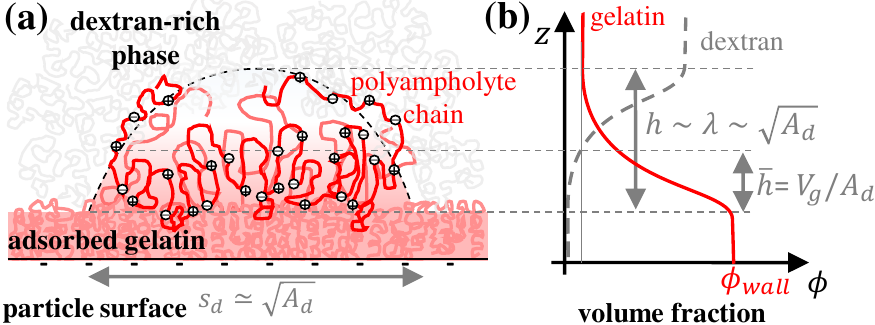}
\vskip -10pt
\caption{Surface defect characteristics.
(a) Adsorbed gelatin morphology. 
(b) Composition profile and characteristic dimensions.}
\label{fig:3}
\vskip -10 pt
\end{figure}

\section{Conclusions \& Discussion}
The characteristic defect dimensions (10--20~nm) that account for our experimental observations are much larger than those previously reported 
(2--4 nm) \cite{kaz2012,rahmani2016} for similar sulfate latex microparticles of radii $R=$~0.5--0.9~$\mu$m at the interface between two molecular liquids. 
To rationalize our findings we first notice that the studied W/W interface with ultra-low surface tension has a rather large effective thickness $2\kappa= k_BT/3 \gamma \xi \simeq 10$~nm \cite{broseta1987,tension2013,tromp2014} and, therefore, surface features smaller than 10 nm cannot induce significant interfacial deformation and pinning.
Furthermore, W/W interfaces have a preferential curvature radius $R_{p}=2\delta+\sqrt{4\delta^2-(3/\gamma)(2k+k_G)}$ that minimizes the excess free energy and is prescribed by the Tolman length $\delta$, and the bending and Gaussian curvature rigidities $k$ and $k_G$, respectively \cite{tension2013,tromp2014}.
For the studied water-in-water emulsion, with both components having similar polymerization degree $N\simeq 1000$, we have $\delta\simeq 0$ and $2k+k_G\simeq k_BT$, which gives $R_{p}\simeq 10$~nm \cite{tension2013} as the preferential radius of gelatin ``globules'' that would nucleate in the dextran  phase.

Total organic carbon content and zeta potentials measurements (see Supplemental Material) indicate that gelatin dispersed in the dextran  phase is adsorbed to the particle surface at the studied polymer and salt concentrations.
As illustrated in Fig.~3, we speculate that gelatin adsorption at the particle surface is mediated by the spreading and adhesion of multiple ``globules'' of volume $V_g=(4/3)\pi R_p^3$, which can produce surface defects with projected areas $A_d\simeq 515$~nm$^2$, mean thickness 
$\bar{h}$ = $V_g/A_d\simeq$~8--10 nm, and tails extending to heights $h$ comparable to the estimated defect sizes $\lambda=\beta \sqrt{A_d}$.   
Neutron scattering studies of various gelatins adsorbed on polystyrene latex particles report a nearly exponential decay of the local volume fraction $\phi(z)$ and a r.m.s. layer thickness between 8 and 10 nm for aqueous solutions at 1~\%w/w and 10 mM salt concentration \cite{cosgrove1998,turner2005}.
Under studied conditions with 25 mM salt concentration and pH=5.8, the Debye length is about 1.9 nm and fish gelatin chains have a weak net positive charge with roughly 10\% of monomers being charged.
A mean thickness $\bar{h}\simeq$ 8--10 nm roughly corresponds to the length of loops that form between charged monomers when the polyampholyte (gelatin) chain (see Fig.~\ref{fig:3}) is adsorbed through combination of dipole-charge and hydrophobic interactions \cite{neyret1995,dobrynin1997,vaynberg2000,zhulina2001}.
According to mean field theory\cite{broseta1987,tension2013,tromp2014} the local volume fraction is $\phi(z)\simeq\phi_{wall}(1+\tanh(z/\kappa))/2$, assuming full coverage at the wall $\phi_{wall}= 1$ the gelatin volume fraction decays to 1\% at a height $h\simeq \sqrt{A_d}=$~22.7 nm from the wall.

Another relevant finding is that the damping coefficient $f_d=6\pi\mu_e R$ scales linearly with the particle radius and shows no dependence on the particle position, even though the particles move from a less viscous dextran  phase to a more viscous gelatin  phase. 
A position-independent damping coefficient is actually consistent with scaling laws for the diffusion coefficient $D=k_B T/f_d=D_o f(\xi/L)$ in semidilute polymer solutions
\cite{degennesscaling,langevin1978,phillies1986,yu2002}, 
where $D_o$ is the diffusion coefficient for pure solvent (zero-concentration), $L$ is the characteristic size of the diffusing ``probe'', and $\xi$ is the blob length or mesh size of the polymer matrix. 
Hence, similar damping coefficients $f_d=k_B T/D_o f(\xi/L)$ are expected in the dextran  and gelatin  phase since both phases have the same zero-concentration diffusion coefficient $D_o$ and similar blob size $\xi$.
Considering $L\sim h \gtrsim$~10 nm is determined by the length of polymer loops and tails extending far to the particle surface (cf. Fig. 3a), then 
$\xi/L\lesssim 1$ and topological hindrance due to entanglement with the surrounding matrix is expected to dominate over hydrodynamic damping \cite{yu2002}.  
It is worth noticing that the scaling of diffusion and damping coefficients with the ratio $\xi/L$ is observed when the polymer matrix is ``quasi-static'' and relaxation mechanisms such as constraint release and contour length fluctuations have characteristic time scales $\tau$ much smaller than the time scale of motion of the ``probe'' \cite{yu2002}. 
The early exponential decay to equilibrium and weak variation of damping coefficients with position for all studied particle radii indicates that $\tau \ll T_D$, where $T_D =$ 55 ms is the shortest decay time observed for $R=0.5$~$\mu$m.
This is consistent with the Newtonian behavior ($\mu=$~const) observed in our rheological measurements for shear rates up to 100 s$^{-1}$. 
We therefore attribute the weak dependence on position of the damping coefficients to the similar topological properties of the studied semidilute solutions.

Experimental results and analysis in this work constitute an initial step toward understanding the complex dynamics of particle adsorption at W/W interfaces in water-in-water emulsions. 
Quantitative agreement produced by the proposed analytical models seems to identify specific mechanisms controlling the observed dynamics and kinetics of adsorption. 
Further work studying colloidal particles with different sizes and surface functionalities at different water-in-water emulsions is needed to further verify the physical assumptions and validity range of the proposed analytical models.  

\begin{acknowledgments}
The authors thank J.H. Snoeider and H.A. Stone for useful discussions. CEC acknowledges partial support from NSF-CBET 1614892.
\end{acknowledgments}

%

%

\end{document}